\documentclass[prl,twocolumn,aps,amssymb,footinbib,showpacs]{revtex4}
\usepackage{bm}
\usepackage{amssymb}
\usepackage{times}
\usepackage{latexsym}
\usepackage{graphicx}
\usepackage{color}
\usepackage{subfigure}

\begin{document}
\title{ Chern and Majorana Modes of Quasiperiodic Systems}
\author{ Indubala I Satija and Gerardo G. Naumis }
\affiliation{School of Physics Astronomy and Computational Sciences, George Mason University,
 Fairfax, VA 22030}
 \affiliation{ Departamento de Fisica-Quimica, Instituto de Fisica, Universidad Nacional Autonoma de Mexico (UNAM), Apdo. Postal 20-364, Mexico D.F., Mexico}
\date{\today}
\begin{abstract}
New types of self-similar states are found in quasiperiodic systems characterized by topological invariants-- the Chern numbers.
We show that the topology introduces a competing length in the self-similar band edge states transforming peaks into doublets of size
equal to the Chern number. This length intertwines with the quasiperiodicity and introduces an intrinsic 
scale, producing Chern-beats and nested regions where the fractal structure becomes smooth.
Cherns also influence the zero-energy mode, that for quasiperiodic  systems which exhibit exponential localization, is related to the {\it ghost} of  the Majorana; the delocalized state at the onset to topological transition.
The Chern and the Majorana,  two distinct types of topological edge modes, exist in quasiperiodic superconducting wires.
\end{abstract}
\pacs{71.23.Ft, 05.30.Rt, 42.70.Qs, 73.43.Nq}
\maketitle

The revelation that quasicrystals belong to topologically nontrivial phases\cite{QCti} of matter is an exciting new development
that opens new avenues in the frontiers of topological insulators. 
Quasicrystals (QC) are fascinating ordered structures exhibiting self-similar properties
and long range order with regular Bragg diffraction\cite{QC}. Characterized by a reciprocal space of dimensionality
higher than the real space, these systems 
do not exhibit crystallographic rotational symmetries.
Topological insulators are exotic states of matter that are insulators in the bulk but conduct along the edges\cite{TI},
characterized by topologically protected gapless boundary modes. 
These edge modes are a manifestation of the nontrivial band structure topology
of the bulk\cite{TKKN} and their number equals\cite{Hat} the topological integer, the Chern number.

Key to the topological characterization of QCs is the
translational invariance that shifts the origin of quasiperiodic order\cite{QCti} and manifests as an
additional degree of freedom relating QCs to higher dimensional periodic systems. 
These shifts known as phasons produce
QCs that look locally different but belong to
the same isomorphism class\cite{Levine}. 
Topological description of QCs requires an ensemble of such systems and can be characterized by Chern numbers in view of their
mapping to higher dimensions. Explicit demonstration of transport, mediated by the edge modes
has been demonstrated by pumping light across photonic QC\cite{QCti,HarperFib,QCphotonic}. These studies have revitalized interest in quasiperiodic systems\cite{Others}, which is a new way to realize topological states of matter.

This paper elucidates a novel manifestation of topology that is unique to QCs.
In 1D QCs, we show that the
{\it band edge modes} encode topological invariants in their spatial profiles.
Characterized by goldenmean incommensurability, the central peaks and the sub-peaks separated by Fibonacci distances
of the band edge states split into doublets of size equal to
the Chern number. 
This Chern-dressing is accompanied by new spatial patterns that  include regions
where the wave function varies smoothly as well as appearance of {\it dimerized} peaks where two consecutive sites have equal amplitudes.
In other words, the topology generates non-fractal local regions of sizes dictated by the Chern, embedded in self-similar structure, reminiscent of
the periodic orbits coexisting with chaotic dynamics.
The Cherns also leave their fingerprints on the momentum distribution of particles, 
the observables that can be measured in experiments involving ultracold gases.
These topological fingerprints in the band edge modes and the momentum distribution
are found in Harper,
Fibonacci as well as in generalized models that interpolate between these two.

QCs also provide a new perspective
on Majorana modes, zero-energy topologically protected modes at the ends of infinitely long superconducting wires with open boundary conditions\cite{Kitaev}.
These modes have been the subject of very intense studies due to their potential applications in quantum computing.
We show that in QCs where quasiperiodicity induces localization, fluctuations about
exponentially localized zero energy modes describe  the onset to a topological phase transition. These delocalized modes, corresponding to the extinction of Majorana ( Majorana ghost) are found to be shadowed by the Cherns.
A perturbed Harper model with broken $U(1)$ symmetry, 
host both the Majorana and the Chern modes.
The Cherns exist in all gaps except the central gap where the Majorana resides.

We consider a 1D chain of spinless fermionic atoms in a lattice 
described by the Hamiltonian,
\begin{equation}
H(\phi)=-\sum_{\bf s} c^\dagger_{\bf s} c_{\bf s+1} + {\rm h.c.} - \sum_{\bf s} V_n(\phi) c^\dagger_{\bf s} c_{\bf s}
\label{harperH}
\end{equation}
Here, $c^\dagger_{\bf s}$ is the creation operator for a fermion at site ${\bf s}$. In our studies we have investigated a generalized potential
that interpolates between the Harper and the Fibonacci model\cite{HarperFib}. However, here we will restrict ourselves to
the Harper model with $V_n=2\lambda \cos( 2\pi (\sigma n+\phi))$,
incommensurate potential
characterized by an irrational number $\sigma$ which we take to be the golden mean ($(\sqrt{5}-1)/2$). Here $\lambda$ controls
the strength of quasiperiodic disorder and $\phi$ is an arbitrary phase.
The Eigenvalue equation, namely the Harper equation,
\begin{equation}
\psi^r_{n+1}+ \psi^r_{n-1} + 2\lambda \cos( 2\pi (\sigma n+\phi))\psi^r_n = E \psi^r_n
\label{Harper}
\end{equation}
exhibits a self-similar spectrum and wave functions at $\lambda=1$\cite{Harper,Ostlund}. This self-dual point is the critical point for
quasiperiodic disorder-induced quantum phase transition
from extended to exponentially localized phase\cite{Harper}.

The incommensurate system is studied by approximating the golden mean, $\sigma$
by a sequence of rational approximates, the ratio
of two consecutive Fibonacci numbers: the
Fibonacci sequence is defined by $F_0=0$, $F_1=F_2=1$, and $F_n=F_{n-1}+F_{n-2}$.
For any rational approximant $\sigma = p/q=F_{n-1}/F_n$, the system consists of $F_n$ bands and $F_n-1$ gaps.
The eigenstates of the incommensurate system are obtained as limiting case of Bloch states of periodic system with period 
$q=F_n$ as $n \rightarrow \infty$, characterized by the Bloch vector $k$.

In parallel with the well known\cite{TKKN} formalism of quantized Hall conductivity for a 2D system, one can define
an {\it adiabatic conductivity} $\sigma_{\Phi}=\frac{e^2}{\hbar} C_r$ for a 1D ensemble $H(\phi)$ of chains with a periodic boundary condition,
$\sigma_{\Phi} = \frac{e^2}{\hbar} \mathrm{Im}\sum_{l=1}^{r} \int d\phi [\int dk \sum_{n=1}^q\partial_{k}(\psi^l_n)^*
\partial_{\phi} \psi^l_n]$.
Here $r$ labels the gap characterized by a Chern number $C_r$ and integration over
$\phi$ corresponds to an ensemble average of set of chains related to each other by translation using phason shifts \cite{Levine}
controlled by $\phi$.

\begin{figure}
\includegraphics[width =1.25\linewidth,height=1.25\linewidth]{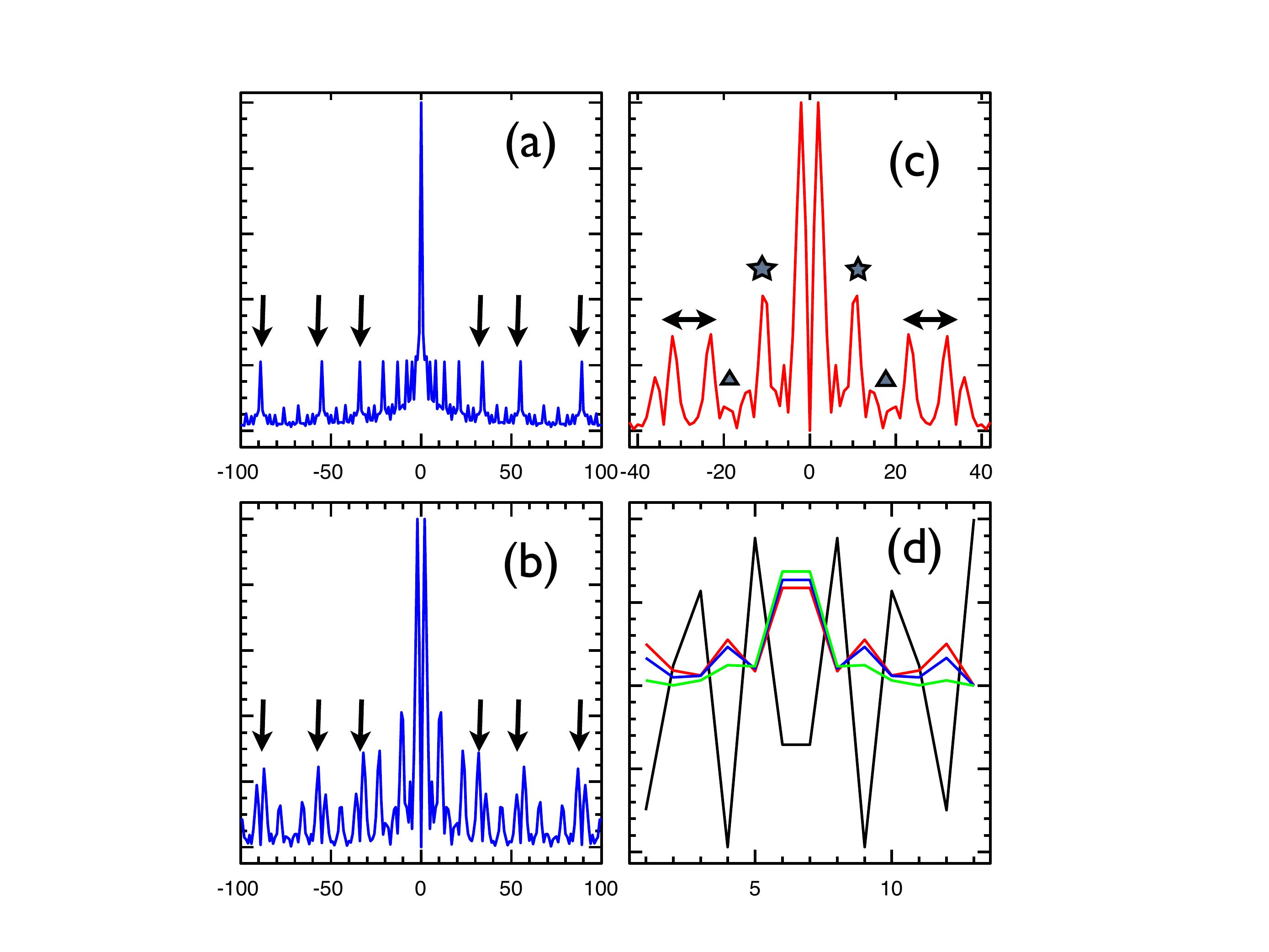}
\leavevmode \caption {(Color online)  Magnitudes of the wave functions for the topologically trivial ground state  (a), and the topologically non-trivial Chern-$4$-- band edge state (b) at $\lambda=1$.
The vertical arrows show the Fibonacci sites which are Chern dressed for the topological states.  Note that the asymmetrically split Fibonacci peaks regain symmetry
as one moves far from the center.
Blowup of the region near the peak for Chern-$4$ state (c), showing novel structures consisting of dimerized peaks (star),
, distorted $M$-shaped regions ( triangle)
and regions separated by $9$-sites (double arrow) where the wave function varies smoothly. 
Panel (d) shows onsite-potential ( black) and the Chern-$1$ wave function
for $\lambda=1, 1.1, 1.5$ ( from bottom-top),  illustrating our key observation that the location of the peaks is independent of $\lambda$.}
\label{wfkall}
\end{figure}

\begin{figure}
\includegraphics[width =1\linewidth,height=1\linewidth]{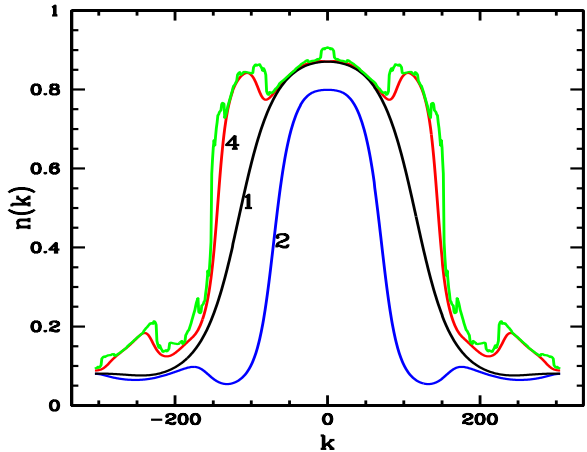}
\leavevmode \caption {(color online) Fingerprints of topological invariants in the momentum distributions for $C_r=1,2,4$
for $\lambda=1$. The  smooth
momentum distribution for the bulk-insulating states develops wiggles. The outermost curve shows the half-filled case with fractal distribution that appears to be strongly
influenced by the Chern-$4$ state which exists in close proximity ( in energy ) to the zero-energy state.}
\label{Mom}
\end{figure}

We now show that topology introduces a new length, equal to the Chern number in the band edge states. 
Fig. \ref{wfkall} shows numerically obtained self-similar wave functions for the band edge modes for topologically trivial and non-trivial states,
using a rational approximant of the golden mean with a periodic boundary condition.
These spatial profiles consist of a central or main peak and secondary peaks at Fibonacci distances 
from the central peak.
The wave functions display self-similarity as the structure around subpeaks approaches a scaled version of the structure around the central peak.
Blowup of the region near any peak reveals new patterns. First, the primary as well as the secondary peaks
split into double-peak structures separated by the distance
equal to the Chern number. This topological encoding of the peaks will be referred as {\it Chern dressing}.
In addition, we see new structures, which are displayed for Chern-$4$ state in Fig. \ref{wfkall}(a-c) and in supplementary material for other states.
We see dimerized peaks where two neighboring sites have almost equal intensity,
a distorted M-shaped structure and a non-Fibonacci peak separation of $9$ where the wave function between the two peaks varies smoothly.
Observed smooth regions  are found to be confined to narrow windows of  sizes $F_n-C_r$ and hence are due to reduction in Fibonnaci spacing between the peaks due to Chern splitting. ( See additional
details in the supplementary section.)
The appearance of local regions of size controlled by the topological invariant
, where the wave functions varies smoothly is rather striking. These smooth pockets signal a very unique
role of topology, namely smoothening of the fractal structures.
These extended structures repeat for every third 
Fibonacci spacing, resulting in a scale invariant wave function as illustrated further in the supplementary material.

We recall that all previous scaling analysis 
of the wave functions for QCs have been carried out for special points of the spectrum such as mid-band points\cite{Ostlund}
or the band edge points corresponding to  maximum or minimum energy states. Such states are topologically trivial.
These studies describe incommensurate states as consisting of
a central or main peak and a sequence of sub peaks at Fibonacci distance from the central peak as shown in Fig.\ref{wfkall}(a). For subpeaks far from the central peak, the ratio of subpeaks to central peak intensity approaches a 
well defined universal ratio  $\zeta$.
Topological states exhibit all these above mentioned features with additional novel structures such as Chern-dressing and smooth regions
embedded in the fractal pattern.

To understand Chern dressing we begin with Thouless'\cite{Thou83} analysis of band edge states
which identifies these wave functions as states of definite parity, with symmetry points about $n=0$ or $n=q/2$. 
In other words, the $r^{th}$ band edge state shows the property, $\psi_{n_1(r)}=\pm \psi_{n_2(r)}$ 
($n_1(r)=-n_2(r)$ 
for center of symmetry about $n=0$ ).
This symmetry  where each site is paired with another site, provides a starting point to understand the Chern dressing. We first consider the limit 
$\lambda \rightarrow \infty$ where
due to exponential localization only the dominant peaks survive.
In this case the
wave function is given by,
$\psi^r_{n}(\pm)=\frac{1}{\sqrt{2}}\left(\delta_{n,n_1(r)}+e^{i\beta_\pm}\delta_{n, n_2(r)}\right)$\cite{fradkin,Zhao}.
It has been shown that the Chern number is given by
$C_r=|n_1(r)-n_2(r)|$, corresponding to the spacing between the two localization centers.
Here
$\beta_{+}=-(C_r-1)\pi $ and $\beta_{-} = -C_r\pi$ 
are relative phases for the upper and lower band edges, respectively. 
We note that the cosine potential also has a pairing property, that is, for every site $m_1$ there exists a site $m_2$ such that $V(m_1)=V(m_2)$.
Since for $\lambda \rightarrow \infty$ the eigenvalues are $E_r(\phi)= -2\lambda \cos( 2\pi \sigma r+\phi)=V(n_1(r))=V(n_2(r))$ and each eigenstate is simultaneously
localized at $n=n_1(r)$ and $n=n_2(r)$,
thus the spacing between the two localization centers is equal to the distance between the paired sites of the potential.
Therefore, as we sort the eigenvalues or the potential, we can associate to each
Chern number $C_r$ a unique pair of sites ( $n_1(r), n_2(r)$ ) of the potential.

We will now argue that the Chern dressing illustrated above for $\lambda \rightarrow \infty$ also occurs for finite $\lambda$
including $\lambda=1$ for the central as well
as for subpeaks.
Our key observation 
( illustrated in  Fig. \ref{wfkall}(d) for Chern-$1$ state) is that the locations of the (local) dominant peaks remain unchanged with $\lambda$.
In other words, for all eigenstates, the
wave function has maximum amplitude at sites
 given by the ordered pair of sites that describe localization centers
for $\lambda \rightarrow \infty$ all the way down to $\lambda=1$ where the system exhibits power-law localization.

The key to hierarchical manifestation of topology is the fact that the symmetry of the band edge wave functions
about $n=0$ or $n=q/2$ as discussed earlier,
manifests asymptotically at all Fibonacci sites and the sites between the two consecutive Fibonacci spaced sites.
This is a direct consequence of the number theory as
$lim_{m\rightarrow \infty}V(F_m)=V(0)=1$ since 
$V(F_m)=2\lambda cos(2\pi \sigma^{m-1})$ and $\sigma^m \rightarrow 0$ for $\sigma<1$.

Chern splitting of the Fibonacci  peaks  introduces a   beat  frequency, which we will refer as {\it Chern-beats}, in  the spatial patterns of the quasiperiodic lattices.
It can be shown ( see supplementary section) that  for the lower band edges, with $\lambda \leq 1$,  $\psi^r_{n}\approx cos(\pi C_r n/q)\Psi^{0}_{n}$, where $\Psi^{0}_{n}$ is the ground state wave function. 
This is the familiar standing wave at the band edges  and for incommensurate lattices has its  period controlled by the Cherns and demonstrates
that the Cherns modulate the existing length scales of  the quasiperiodic system.
A length scale rooted in topology may underlie a whole class of physical phenomena, such as
{\it mode beating} observed in quasiperiodic systems\cite{beat}.

The topological fingerprints are also encoded in the momentum distribution
, $n(k) = \sum_m |\eta^k_m|^2$ where $\eta^k_m$ is the
Fourier transform of the wave function $\psi_n$.
The measurement of these variables is routinely done in time of flight experiments in ultracold atomic gases.
As shown in the figure
~(\ref{Mom}), the smooth momentum distribution, characteristic of the insulating state,
develops wiggles that are correlated with the
topological invariant. In view of self-duality, the momentum distribution also
describes the density profile with $k=<\sigma n>$.  It should be noted that the fermionic density also describes the density for a gas of hard core
bosons which have been realized in cold atomic gases\cite{HCB}. 

\begin{figure}
\includegraphics[width =1.1\linewidth,height=1\linewidth]{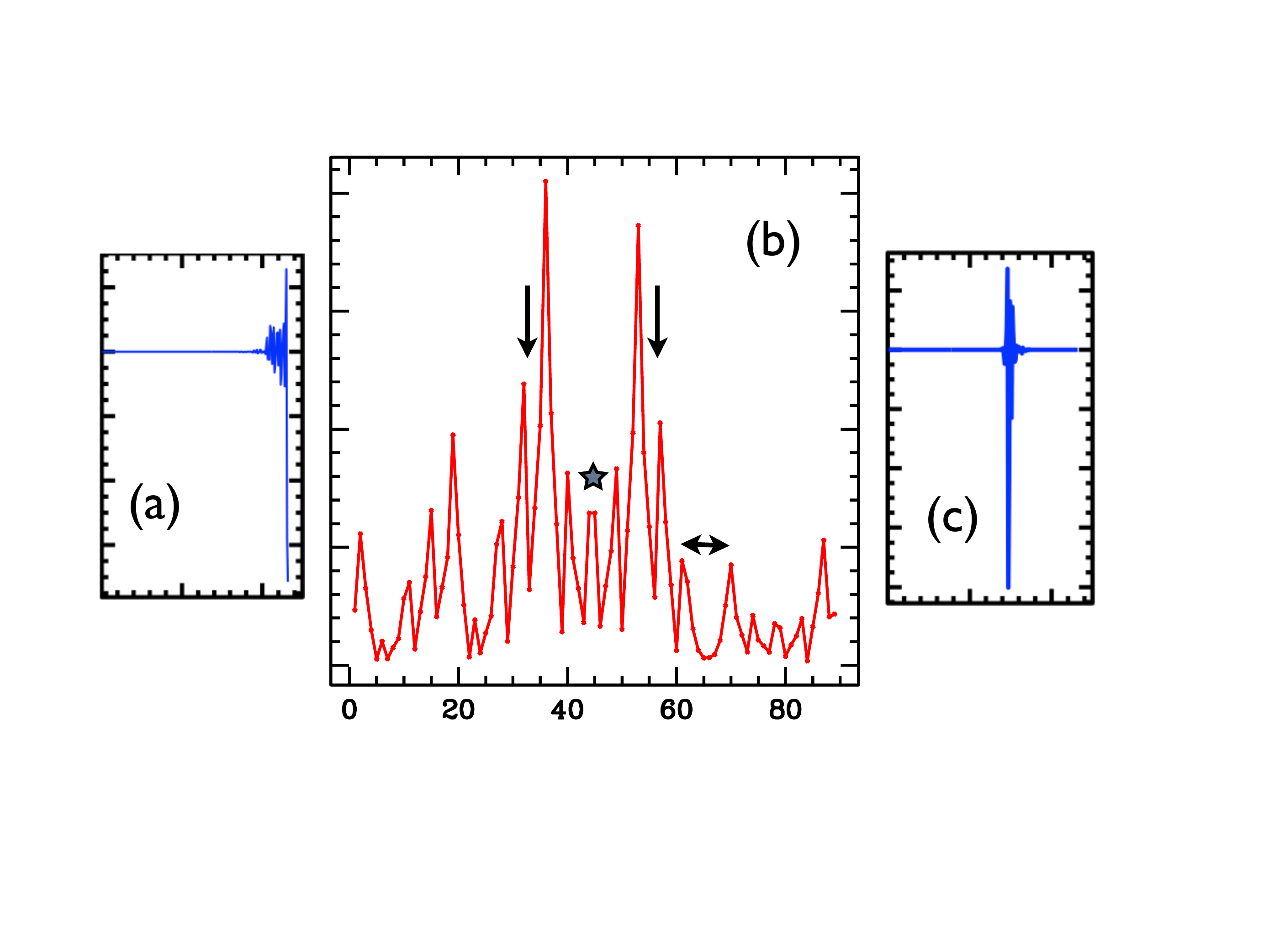}
\leavevmode \caption {(color online) With $\Delta=.02$, panels a--c respectively show the magnitude of
the wave function below, at and above the topological phase transition. Stars and double-arrows identify regions that resemble the
Chern-$4$  state shown in Fig.~(\ref{wfkall}).}
\label{Major}
\end{figure}

We now discuss  quasiperiodic systems that support Majorana modes\cite{Kitaev} in addition to the Cherns discussed above.
We first note that in QCs such as the Harper model exhibiting exponential localization, fluctuations about
exponentially decaying envelope of zero energy state are related to the zero energy excitations of a superconducting wire\cite{KS}.
From Eq. (\ref{Harper}), it follows that the
fluctuations $\eta_n$ about
exponentially decaying envelope, $\psi_n = e^{-n\xi} \eta_n$ satisfy the following equation,
\begin{equation}
e^{-\xi} \eta_{n+1} + e^ {\xi} \eta_{n-1} + 2 \lambda \cos( 2\pi (\sigma + \phi)) \eta_n = E \eta_n
\label{fluc}
\end{equation}
where the localization length $\xi^{-1}= ln \lambda$.
For $E=0$, this equation describes a fermionic representation of the zero energy state of a
spin-$1/2$ anisotropic XY-chain with exchange 
interactions $J_x=e^{-\xi}$ and $J_y =e^{\xi}$ along the $x$ and the $y$ axis respectively,
in a spatially modulated transverse magnetic field $V_n(\phi)$\cite{KS}. 
The system also describes a
$p$-wave superconducting quantum wire with superconducting gap parameter $\Delta=J_y-J_x = \lambda-1/\lambda$. 
The eigenvalue equation of the system is a coupled set of equations,
\begin{eqnarray}
J_x f_{n+1}+J_y f_{n-1} + V_n(\phi) f_n & = & E g_n\\
\label{sup1}
J_y g_{n+1}+J_x g_{n-1} + V_n(\phi) g_n & = & E f_n
\label{superc}
\end{eqnarray}
Here $(f_n, g_n)$ represents a two-component wave function of the superconducting chain with particle-hole symmetry. 
For $E= 0$, eqns.(\ref{fluc}), (\ref{sup1}) and (\ref{superc}) coincide.
For $\Delta \ne 0$, this perturbed Harper with broken
$U(1)$ symmetry ( where fermion number is not conserved ) has been shown to exhibit a localization transition at the critical
value $\lambda_c=J_y= e^{\xi}=\lambda$,
beyond which all states are exponentially localized\cite{KS}. 
Therefore, the {\it zero energy localized states} of Harper are related to the zero energy states of superconducting chain
{\it only at the onset of topological phase transition} where the superconducting model becomes gapless.

In a finite chain, the gap at $E=0$ supports Majorana fermions, which are edge localized  modes for $\lambda < \lambda_c$.
To study  Majorana modes in the interior of the topological phase, $U(1)$ symmetry breaking
superconducting model ( Eq. (\ref{superc})) is used. Here we confine ourselves mostly to small $\Delta$
values which leave the Chern modes of the Harper model almost unchanged and in addition support the Majorana modes at $E=0$.
Below criticality ( $\lambda < \lambda_c$ ), superconducting model is gapped at $E=0$, making it topologically distinct from the Harper equation.
The  Chern modes exist in all gaps
except the central gap that hosts the Majorana. Figure (\ref{Major}) shows the edge localized Majorana in the topological phase, its delocalization at the critical point ( which we refer as Majorana ghost) as well as  the mode localized in the interior of the chain where it loses topological protection.
At the onset to the topological transition, the Majorana ghost shows some features that are common to the Chern-$4$ state.
 In quasiperiodic systems with fragmented spectrum,  the smallest  Chern ( and hence most important) state that appears close to zero-energy state is the Chern-$4$ state  and therefore we see its
 influence or {\it shadows}  on the Majorana ghost, seen clearly in its spatial profile ( Fig. ~\ref{Major}) and also in the momentum distribution (Fig. ~\ref{Mom}).

In summary, the band edge states of the quasiperiodic systems exhibit  new types of self-similar patterns
as the topological length intrudes the fractal self-similar structure. Topology also dresses peaks with Cherns and smoothers some parts of the fractal structure. This is 
reminiscent of ordered structures, such as periodic orbits , embedded in a chaotic sea in Hamiltonian systems.
Interplay between topology and quasiperiodicity provides a remarkable example of competing periodicities
where Cherns, the aliens, adapt to the landscape controlled by the Fibonacci length scales.
Our results are valid for all irrationals with a periodic tail in the continued fraction expansion\cite{Ostlund}.
The renormalization scheme that incorporates the Chern length, preserving scale invariance, remains an open problem. 

In addition to photonic systems, QCs described by the Harper equation have been realized in 
ultracold atomic gases\cite{twocolor}.  From the cold-atom points of view, these incommensurate lattices have been investigated in various contexts\cite{QP}.
Superconducting quantum wires have been shown to be promising candidates
for realizing Majorana modes\cite{SC}.
Quasicrystalline superconducting chains offer a fascinating
new set of possibilities to study topological states supporting both the Majorana and the Cherns.

This research is supported by ONR and DGAPA-UNAM IN102513.

\end{document}